# Effect of Oxygen Content on the Properties of the HoBaCo$_2$O$_{5+\delta}$ Layered Cobaltite

Yuri Fernandez-Diaz[1], Lorenzo Malavasi[1,*], Maria Cristina Mozzati[2]

[1]Dipartimento di Chimica Fisica "M. Rolla", INSTM and IENI-CNR, Università di Pavia, Viale Taramelli 16, 27100 Pavia, Italy.

[2]CNISM, Unità di Pavia and Dipartimento di Fisica "A. Volta", Università di Pavia, Via Bassi 6, I-27100, Pavia, Italy.

*Corresponding Author: Dr. Lorenzo Malavasi, Dipartimento di Chimica Fisica "M. Rolla", INSTM, Università di Pavia, V.le Taramelli 16, I-27100, Pavia, Italy. Tel: +39-(0)382-987921 - Fax: +39-(0)382-987575 - E-mail: lorenzo.malavasi@unipv.it

# Abstract

In this work we present the result of the first experimental investigation of the role of oxygen content on the properties of the HoBaCo$_2$O$_{5+\delta}$ layered cobaltite. We have measured the variation of oxygen content as a function of temperature and oxygen partial pressure by means of thermogravimetry coupled to chemical titration analysis. On selected samples of accurately known oxygen content we have undertaken a systematic investigation of their structural, thermal and magnetic properties by means of x-ray diffraction, differential scanning calorimetry and magnetometry. The overall results gained by this study confirm the central role of oxygen content on the properties of these materials suggesting that, for the Ho composition, even very slight variation of the order of δ=0.01 has a dramatic influence on the magnetic and transport properties of the samples. In addition, we have presented results showing the strategy to check the quality of samples prepared at selected oxygen contents by annealing procedures.

# Introduction

Recently there has been a growing interest in the study of layered cobaltites of general formula $REBaCo_2O_{5+\delta}$ ($RE$=rare earth) due to their rich structural, electronic, and magnetic phase diagrams resulting from the strong coupling between charge, orbital, and spin degrees of freedom [1-5]. These strongly correlated electron systems have attracted a great deal of attention as a promising alternative to conventional semiconductors in the field of thermoelectric power generation [6].

One of the most interesting aspect of layered cobaltites is the large oxygen non-stoichiometry they show, whit $\delta$ theoretically varying from 0 to 1. This degree of freedom allows a continuous doping of the square-lattice $CoO_2$ planes which not only influences the mean Co valence state but also the carrier nature, the Co spin-state and the bandwidth. For example, considering the $REBaCo_2O_{5.5}$ parent compound with all cobalt ions in the +3 valence state the change in the oxygen content around this stoichiometry allows a doping with both electrons ($Co^{2+}$ ions) or holes ($Co^{4+}$ ions) [7]. Figure 1 shows a sketch of the orthorhombic crystal structure of a generic $REBaCo_2O_{5.5}$ cobaltite. As can be seen, this composition is characterized by an equal numbers of ordered $CoO_6$ octahedra and $CoO_5$ square pyramids. Upon changing the oxygen content, some oxygen ions are inserted into or removed from the $REO_x$ planes, which changes the numbers of $CoO_6$ octahedra and $CoO_5$ pyramids and also creates electrons or holes in $CoO_2$ planes.

Among the most interesting properties of layered cobaltites, which are tuned by the oxygen content, Taskin *et al*. have shown a remarkable change in the charge carriers nature in $GdBaCo_2O_{5+\delta}$ and $NdBaCo_2O_{5.\delta}$ single crystals moving from $\delta \approx 0.40$ to $0.60$ allowing them to give a solid experimental support to the idea that strong electron correlations and spin-orbital degeneracy can bring about a large thermoelectric power in transition-metal oxides [8]. This feature was first observed by Maignan *et al*. on the $HoBaCo_2O_{5.5}$ composition where they proposed a model, to explain the change in the transport properties, based on a complete or partial conversion of the HS $Co^{3+}$ located in the octahedra to the LS state. This conversion ''immobilizes'' the electron charge carriers due to the phenomenon of spin blockade, and the latter is replaced by an activated regime for holes (LS $Co^{4+}$) moving in the much narrower $t_{2g}$ band [4].

A study of the physical properties of layered cobaltites as a function of the oxygen content may revel interesting and new features on their properties. However this kind of study is experimentally quite demanding particularly in systems where fast oxygen diffusion occurs, as in the present case. As a matter of fact only one thorough investigation of this aspect has appeared in the current literature on the $GdBaCo_2O_{5+\delta}$ composition [7]. For other rare earths the available works report data on a limited number of samples as a function of $\delta$ as for the $PrBaCo_2O_{5+\delta}$ [9] and $NdBaCo_2O_{5+\delta}$ [10]. However, this last work contains quite questionable data where most of the samples investigated are bi-phasic in nature as a result of a poor control of the oxygen content and the optimization of the procedure needed in order to obtain high-quality samples.

In this paper we undertook a study of the role of oxygen content on the physical properties of the $HoBaCo_2O_{5+\delta}$ sample. This particular rare earth has been chosen since it was the composition showing the most impressive resistivity change at the I-M transition (located at about 305 K). In addition Ho is the smallest rare earth and thus, to carry out a systematic investigation of the $REBaCo_2O_{5+\delta}$ family, we decided to start from this "limit" of the various $RE$ present in the $REBaCo_2O_{5+\delta}$ lattice.

In the present work we are going to give experimental results regarding the tuning of the oxygen content in this composition and also show the experimental difficulties behind the control of this variable. On several samples with varying oxygen content we carried out a structural and magnetic characterization by means of X-ray diffraction and SQUID magnetic measurements. In addition, the presence of I-M transitions has been checked by means of differential scanning calorimetry (DSC).

# Experimental Section

Powder samples of $HoBaCo_2O_{5+\delta}$ have been prepared by conventional solid state reaction from the proper stoichiometric amounts of $Ho_2O_3$, $Co_3O_4$, and $BaCO_3$ (all Aldrich ≥99.99%) by repeated grinding and firing for 24 h at 1050-1080 °C. $Ho_2O_3$ was first heated at 900°C overnight before being used in the reaction.

Oxygen content was fixed according to thermogravimetry (TGA) measurements by annealing $HoBaCo_2O_{5+\delta}$ pellets at selected $T$ and $p(O_2)$ in a home-made apparatus for at least 48 hours followed by rapid quenching in liquid nitrogen. This assures a very limited oxygen in-diffusion during the cooling step and a rigorous control of the oxygen content. This procedure allowed us to prepare samples with $0<\delta\leq 0.5$. In order to prepare a sample with $\delta=0$ we annealed a $HoBaCo_2O_{5+\delta}$ pellet under high vacuum ($10^{-9}$ bar) at 700°C followed by slow cooling at room temperature at 0.5°C/min. Finally, we also annealed one sample under 100 bar oxygen pressure at 600°C in order to prepare a sample with $\delta>0.5$.

The oxygen content of the samples was determined by means of a well-known iodometric method described elsewhere [11]. In order to avoid the interference of atmospheric oxygen, we evacuated the container from air before dissolving the samples in hydrochloric acid under nitrogen. The nitrogen flux was maintained during the whole titration, since we observed that the end-point was reversible in the presence of air. For each sample the oxygen content reported is the average of at least three titrations, with standard deviation below 1%. In order to avoid any Co oxidation during the titration we evacuated all the solution from air and we carried out the procedure under a dry argon atmosphere.

X-ray diffraction (XRD) patterns at room temperature were acquired on a "Bruker D8 Advanced" diffractometer equipped with a Cu anode in a $\theta$-$\theta$ geometry. Measurements were carried out in the angular range from 10 to 110° with 0.02° step-size and acquisition time for each step of at least 10 s. Diffraction patterns were refined by means of Rietveld method [12,13] with the FULLPROF software [14]. Sample chemical composition was checked by means of Electron Microprobe Analysis (EMPA) and it was found in agreement with the expected nominal one.

Static magnetization was measured at 100 Oe from 360 K down to 2 K with a SQUID magnetometer (Quantum Design).

DSC measurements were carried out in static atmosphere using a DT2029 calorimeter. Samples were first equilibrateed at 220K and then slowly heated up to 320K with a heating rate of 2°C/min.

Thermogravimetric measurements were used to determine the oxygen content. These measurements have been performed under different atmospheres, namely $p(O_2)$=1, $10^{-2}$, $10^{-4}$ and $10^{-6}$ atm from 473 to 973 K with a TA 2905 thermal analysis system.

# Results and Discussion

*1. Oxygen Content*

Figure 2 shows a typical TGA trace of HoBaCo$_2$O$_{5+\delta}$ as a function of temperature. In particular, the measurement displayed in Figure 2 refers to the weight change in pure oxygen. As can be appreciated the weight variation is significant and the behaviour of the weight change is highly reversible.

From this kind of measurements and taking different "fixed" points through the chemical titration we calculated the oxygen content as a function of *T* and *p*(O$_2$) variables which is shown in Figure 3. The minimum oxygen content attainable at ambient pressure, within the *T* and *p*(O$_2$) ranges employed in the present work, is 5.01 and the maximum one is 5.50. We will show later that even at the highest temperature and lower *p*(O$_2$) the samples obtained are not in the average +2.5 oxidation state (HoBaCo$_2$O$_5$) and, as mentioned in the Experimental section, this sample had to be prepared by high-temperature annealing in high vacuum. The sample annealed under 100 bar oxygen pressure resulted to have an oxygen content of about 5.540. This result suggests that even under a highly oxidizing treatment the HoBaCo$_2$O$_{5+\delta}$ compound can not attain high oxygen contents. Looking at the current literature it is possible to notice that high oxygen contents around 6 have been found for the LaBaCo$_2$O$_{5+\delta}$ system while for the Nd and Gd substituted cobaltites δ-values up to 0.75 have been obtained by means of thermal treatments under high pressure analogous to that performed in the present work [9]. It is clear a correlation between the maximum oxygen content attainable and the size of the *RE*. Holmium is the smallest among the rare-earths investigated so far and its behavior related to the amount of oxygen that can be introduced in the lattice agrees well with the dependence with the *RE*-size.

We remark here that the definition of the equilibrium oxygen content for this and related systems is a very delicate procedure which requires an intensive research effort particularly in order to obtain reliable results. However, this piece of information is essential in order to understand the physical properties of the layered cobaltites and to carry out a systematic investigation of the correlation between oxygen content and the physical properties.

Starting from the compound with five oxygen atoms in the structure, the increase of oxygen content (5+δ) can be written according to the following quasi-chemical equilibrium (in the Kroeger-Vink notation):

$$\tfrac{1}{2} O_2 \Leftrightarrow O_i'' + 2h^\bullet \tag{1}$$

the electronic holes created are responsible for the oxidation of the $Co^{2+}$ ions to $Co^{3+}$:

$$Co_{Co}^{x} + h^\bullet \Leftrightarrow Co_{Co}^{\bullet} \tag{2}$$

or, for δ values greater than 5.5, also to the creation of $Co^{4+}$ ions.

This means that the fraction of $Co^{3+}$ ions over the total Co is:

$$\frac{[Co^{3+}]}{[Co]} = \frac{0.5+\delta}{1} \tag{3}$$

So, by tuning the oxygen content it is possible to finely tune the Co valence state. We stress here that the in-diffusion of the oxygen within the lattice may not be random. This is the common situation found at δ=0.5 where there is a regular alternation of $CoO_5$ pyramids and $CoO_6$ octahedra (see Figure 1) which is due to a preferential occupation of one of the two O-sites located at $z=0.5$ in the *Pmmm* structure. Oxygen atoms ordering is however possible also at δ values lower than 0.5 [9].

## 2. X-ray Diffraction

X-ray diffraction and Rietveld analysis have been used to extract information about the structural evolution of the HoBaCo$_2$O$_{5+\delta}$ layered cobaltite as a function of δ. Figure 4 shows a typical refined pattern of a tetragonal sample of HoBaCo$_2$O$_{5+\delta}$. In particular, the Figure refers to the sample with δ=0.22. From a structural point of view these compounds can be described as ordered oxygen-deficient perovskites derived from the simple perovskite by doubling along the *c*-axis, and characterized by 1:1 ordering of the Ba$^{2+}$ and Ho$^{3+}$ cations in the form of alternating planes (see Figure 1) with oxygen vacancies located at the level of the Ho$^{3+}$ layers [15,16].

Let us start to consider the evolution of the crystal symmetry at room temperature along with the increase in the oxygen content from δ=0 to δ=0.55. At δ=0 the crystal structure was refined according to an orthorhombic unit cell (space group, *Pmmm*) with $a\sim b\sim a_p$ (where $a_p$ represents the pseudo-cubic lattice parameter of the perovskite unit cell) and $c\sim 2a_p$. Table 1 reports the structural parameters for the various samples investigated in the present work. At δ=0 the number of Co$^{2+}$ and Co$^{3+}$ ions is equal and all of these cobalt ions are coordinated in corner shared square base pyramids formed by the oxygen neighbors while the La and Ba atoms are ordered and form alternated layers along the *c*-axis. The orthorhombic distortion of the unit cell is extremely small as can be appreciated from the values of the lattice parameters reported in Table 1. However, a clear indication of this distortion was observed in some diffractions such as the (020) and (200) located at about 46°. The presence of an orthorhombic symmetry for the HoBaCo$_2$O$_5$ sample agrees with the general behavior of the *RE*BaCo$_2$O compounds which have been found to present all a slight orthorhombic distortion at room temperature [17]. Interestingly, by even slightly moving away from δ=0 to δ=0.01 (*i.e* by oxidizing 1% of Co$^{2+}$ ions), the structure already becomes tetragonal and the peculiar features of the HoBaCo$_2$O$_5$ sample disappear (see later in the text).

From δ=0.01 to 0.39 all the samples present a tetragonal symmetry (space group, *P4/mmm*) with $a=b=a_p$ and $c\sim 2a_p$. The evolution of the unit cell parameters and cell volume are reported, respectively, in Figures 5 and 6. In this composition range the oxygen ions start to populate the HoO$_\delta$ layer by distributing on the 1*b* Wycoff position that is the (0,0,½) position. The distribution of oxygen can be random or ordered. However, by x-ray diffraction this difference can not be observed since this involves the rise of ordering extra-peaks which can be only revealed by neutron diffraction.

At δ=0.5 the crystal structure can not be longer described within the tetragonal *P4/mmm* symmetry. As can be appreciated from Figure 7 several single diffraction peaks for δ<0.5 samples are now split at δ=0.50. The new symmetry at δ=0.50 is orthorhombic (*Pmmm*) with a doubling (with respect to the tetragonal lattice) of the unit cell along the *b*-axis giving rise to the $a_p \times 2a_p \times 2a_p$ unit cell. The indexed pattern shows in fact that all of the (*h0l*) diffractions are now described by two distinct diffraction peaks. In this sample all the cobalt ions are present in the +3 oxidation state and there is an ordering of the oxygen ions within the HoO$_\delta$ layer with a preferential occupation of the (0,½, ½) with respect to the (0,0, ½) which results in the regular alternation of CoO$_6$ octahedra and CoO$_5$ square base pyramids (see Figure 1). It should be noticed that the crystal structure observed at δ=0.50 most probably exists in a very narrow δ range below this value of oxygen content and requires that all the Co ions are in the +3 oxidation state. This is witnessed by the fact that even a sample with δ=0.45 presents a tetragonal symmetry with a small fraction (about 10% as determined from the Rietveld refinement) of a secondary orthorhombic phase due to a possible slight oxygen content un-homogeneity. However this behavior is indicative that at δ=0.45 we are most probably very close to a boundary for the existence of the orthorhombic symmetry. This result seems to be peculiar for the Ho containing layered cobaltites since for the only other two members (Sm and Eu) for which some samples around δ=0.5 have been inspected the crystal structure has been found to be orthorhombic [18]. However, since no structural data have been shown in this work it is not possible to judge if the samples were monophasic or contained a major orthorhombic phase with δ=0.50 and a secondary phase with a different oxygen content. Looking at the very broad and asymmetric susceptibility curves reported by the Authors of ref. [18], particularly when the deviation from δ=0.50 is significant, the hypothesis that oxygen un-homogenous samples have been obtained may not be discarded.

Finally, the same orthorhombic $a_p \times 2a_p \times 2a_p$ unit cell found at δ=0.50 was found for the sample with δ=0.55. The extra oxygen introduced in the lattice starts to populate the (0,0, ½) position of the $HoO_\delta$ layer thus increasing the number of $CoO_6$ octahedra and leading to a mixed Co valence greater than 3.

Looking at Figure 5 it is possible to observe a progressive reduction of the cell volume as the oxygen content increases up to δ<0.50. The contraction of the cell volume is due to the oxidation of the $Co^{2+}$ ions to $Co^{3+}$ ions since, for the same coordination, the first one has a ionic radius of 0.885 Å (high-spin configuration) while the second has a ionic radius of 0.75 Å (high-spin configuration). However, the cell volume trend as a function of δ is not linear due to the fact that the ions are not always in the same coordination and that there are also changes in the $Co^{3+}$ spin-state as a function of the oxygen content (see later in the text). A quite surprising rise of the cell volume is observed at δ=0.50. This anomalous result has been checked by preparing several samples with this composition and in all the cases we found the same result. The source of the slight sudden expansion of the unit cell at this oxygen content is related to the ordering process of the oxygen ions (and consequently of oxygen vacancies), to the ordered distribution of the Co ions and to the set up of the magnetic ordering which occurs around room-temperature with a jump in the cell volume of about 0.15% which, however, does not account for the whole increase observed here. This jump in the cell volume at δ=0.50 is also confirmed by the structural data above δ=0.50. The cell volume of the δ=0.55 sample decreases with respect to δ=0.50 as a consequence of the Co oxidation. Interestingly, the change in the lattice parameters from δ=0.50 to 0.55 is not isotropic. The lattice constant *a* is the only one which shrinks while both *b* and *c* parameters enlarges.

Concerning the lattice parameters reported in Figure 6, it is possible to note that, for the tetragonal samples, there is a slight reduction of the tetragonal distortion along with the increase in the oxygen content, particularly at low δ values. The trend of the Co-O bond lengths, which is shown in Figure 8, shows in fact a progressive approach of the Co-O1 and Co-O2 lengths (the apical bonds) as δ increases. Co-O2 increases with δ since more and more oxygen ions are populating the HoO$_\delta$ layer as the oxygen content increases while the Co-O1 bond length trend mainly reflects the cell contraction. The average value of the Co-O bonds is reported for all the samples in Table 1.

For the sample with δ=0.50 the orthorhombic structure with the doubling around the *b*-direction leads to two different Wycoff positions for the Co ions: 2*r* at (0,½,*z*) and a 2*q* at (0,0,*z*) with the two *z*-values slightly different for the two sites (0.2397(15) for the first one and 0.2498(17) for the second one). The difference results from the coordination of the Co ions. Those in the 2*q* position are octahedrally coordinated while those in the 2*r* position are coordinated by square base pyramids. It is interesting to note that the distortion of the CoO$_6$ octahedron is mainly within the plane since the Co-O1 and Co-O3 lengths (the two apical bonds) are very close each others (Co-O1=1.8769(13) and Co-O2=1.8790(13)). A sketch of the Co$_6$-CO$_5$ polyhedra with the bond lengths is reported in Figure 9. The inter-octahedra Co-O-Co bond angles is about 168°, the inter-pyramids angle is 155.9 while the inter-polyhedron bond angle is around 162.3°. These values are in perfect agreement with neutron diffraction results recently reported on the HoBaCo$_2$O$_{5.5}$ sample indicating that we are already in the insulating and charge ordered phase, in agreement with the I-M transition temperature for this sample (see Figure 11).

For the sample with δ=0.55 the geometry around the Co ions changes significantly with respect to δ=0.50. As an example, the two apical bonds of the octahedra which were very close each other at δ=0.50 are now different: Co-O1=1.8055(13) and Co-O2=1.9560(13). Overall, the bond lengths and bond angles closely resembles the behavior of the HoBaCo$_2$O$_{5.5}$ sample above the IM transition, i.e. where no orbital ordering is present. The magnetic, transport and DSC data of the HoBaCo$_2$O$_{5.55}$ sample (see later in the text) reveal that a step like increase of the electrical resistivity occurs at a temperature lower with respect to that of the δ=0.50 sample (inflection point in the resistivity curve is around 282

K). The anisotropic change in the lattice constants from δ=0.50 to 0.55 is totally analogous to the change which occurs in the δ=0.50 sample when moving to high temperature from the orbital ordered phase (i.e. from the insulating to the metal phase). To summarize, the structural data coupled to the physical properties of the δ=0.55 sample indicate that at room temperature we are within a metal-like phase (even though the electrical resistivity is one order of magnitude higher than the metallic value at δ=0.50) without any sign of orbital order. The geometrical data also suggest that the lower electrical conductivity is mainly due to the reduction of the inter- and intra-polyhedron bond angles which reduce the electron transfer integral through the Co-O-Co network.

With reference to this last aspect, let us consider the in-plane Co-O-Co bond-angle (φ). This parameter has a quite relevant importance for the properties of the $HoBaCo_2O_{5+\delta}$ layered cobaltite, particularly for the transport properties in an analogous way as found in the manganites [19,20], where the electron transfer integral through the Co-O-Co network is proportional to ~cos(π-φ). Figure 10 shows the trend of the inter-polyhedron Co-O-Co bond angle as a function of δ. As can be seen there is a progressive increase of its value as δ increases suggesting a progressive delocalization of the charge carriers as the oxygen content increases up to δ=0.50.

## 3. Differential Scanning Calorimetry (DSC) and Electrical Conductivity

Differential scanning calorimetry (DSC) has been used to study the possible presence of insulating to metal (I-M) transitions and structural transitions. Previous works have shown that clear DSC peaks are observed at the I-M transition and also at the magnetic transitions [21,22]. It is already known that the sample with δ=0.50 shows a sudden rise of the electrical resistivity at about 300-310 K [4]. The nature of this I-M transition is closely related to the $Co^{3+}$ spin-state: in the metallic regime the carriers are delocalized in a $e_g$ conduction band of the IS/HS $Co^{3+}$ orbitals. At the I-M transition there is a partial or total conversion of the $Co^{3+}$ located in the octahedra to the LS spin state with a concomitant immobilization of the electron charge carriers due to the phenomenon of the spin blockade [4].

Figure 11 shows the DSC curves for some selected sample among those studied in the present work. The only samples which showed endothermic peaks in the DSC curves (done on heating 5°C/min) are the $HoBaCo_2O_5$, $HoBaCo_2O_{5.5}$ and $HoBaCo_2O_{5.55}$ samples. We remark that DSC is a very useful tool since it allows also to detect the possible presence of two phases with close oxygen contents which may give apparent single-phase x-ray diffraction patterns (the two phases have similar lattice constants and only an accurate evaluation of the peak width may reveal the "distribution" of lattice parameters) and one apparent magnetic transition (when the two magnetic transitions are very close just one relatively broad peak is present). As a matter of fact, the red line in Figure 11 (panel A), shows the DSC curve of a $HoBaCo_2O_{5+\delta}$ sample annealed at 700°C in pure oxygen and slowly cooled down to room temperature. This procedure is often applied in the current literature in order to give samples with $\delta=0.50$. As can be seen, even though the XRD pattern – without a thorough analysis performed on samples with varying $\delta$ – may suggest a single-phase material, the presence of two peaks is suggestive of a two-phase composition. In fact through a deep Rietveld analysis, these and other samples which have not been prepared by annealing and quenching in liquid nitrogen, are actually two-phases materials. This situation occurs particularly for treatment under oxygen, where the $\delta$-variation with $T$ is significant (see Figure 2) and the oxygen diffusion is very fast and has a strong dependence with the temperature. On the opposite, the $HoBaCo_2O_{5.5}$ sample prepared by annealing and quenching in liquid nitrogen displays one sharp and narrow transition at about 303 K which perfectly agrees with recently published work on very high quality samples [23]; this I-M transition is clearly revealed by four probes conductivity measurements (shown in the inset of Figure 11, panel A). Only "optimally" doped samples have a sharp I-M transition in this temperature range, while oxygen un-homogeneous samples usually present only slope changes and not the peculiar step rise increase of resistivity shown by these compounds at $\delta=0.50$. This I-M transitions occur concomitantly with the melting of the orbital order in pyramids and increase of the Co-O-Co bond angle together with unit cell volume collapse [23].

The $HoBaCo_2O_{5.55}$ sample has as well a clear endothermic DSC peak at a lower temperature with respect to $\delta=0.50$ (peak max. at 295 K instead of at 303 K). The peak maximum corresponds well to the rise of the electrical conductivity (see the blue curve in the inset of Figure 11 panel A). The transition in the resistivity curve is not sharp as for the $\delta=0.50$ sample and extends from about 295 to 275 K. The width of the I-M transition agrees well with the width of the DSC peak. Let finally note that the conductivity before the I-M transition is lower for the more oxygenated sample but is higher in the insulating phase in all the temperature range explored.

All the other samples in the $0<\delta<0.5$ oxygen content range do not display any DSC peak in the *T*-interval explored and, in addition, have an activated transport regime as a function of temperature (not shown). This is an interesting result since it shows that the possibility of an electronic delocalization occurs only at $\delta=0.50$ where all the Co ions are in the +3 valence state and the oxygen vacancies are well ordered.

The other sample showing peaks in the DSC curve is the one at $\delta=0$ (Figure 11, panel B). In this case there is a first peak at about 345 K and a second one at about 219 K. These two DSC peaks correspond to the on-set of the charge-ordering (CO) of $Co^{2+}$ and $Co^{3+}$ ions and to the Néel temperature ($T_N$) of the antiferromagnetic transition [16,17]. Let us note also in this case that even a very small deviation of the oxygen content from 0 leads to the suppression of both transitions. A sample with $\delta=0.01$ which we have studied in this work does not display any peak in the DSC curve. Again, this technique is a valuable tool to put in evidence transitions in the layered cobaltites and also a useful guide to probe the quality of the samples investigated.

*4. Magnetic Properties*

Figure 12a and 12b reports the magnetic susceptibility ($\chi_{mol}$) and the inverse magnetic susceptibility ($1/\chi_{mol}$) for the samples investigated in the present work. For the samples with $\delta$ greater than ~0.4 a clear ferromagnetic behavior is present followed by its disappearance, which is particularly sudden for the $\delta=0.50$ sample. For samples with $\delta$ lower than ~0.4 an overall paramagnetic behavior is observed.

The low-$T$ magnetic structure of the HoBaCo$_2$O$_{5.5}$ sample is quite complicated and carefully investigated only recently [24]. From this work it has been observed an antiferromagnetic structure with a 2$a_p$×2$a_p$×4$a_p$ magnetic unit cell containing four crystallographically independent Co ions, two octahedrally coordinated and two pyramidally coordinated. Of the two Co ions in the octahedra, one has been found to be in the high-spin (HS) state while the other in a mixed intermediate- (IS) and low-spin (LS) state. The pyramidally coordinate Co$^{3+}$ ions were found to be in the intermediate spin-state [24]. It was also observed that the complex magnetic structure of HoBaCo$_2$O$_{5.5}$ contains both positive and negative exchange interactions between nearest neighbors. The region of the paramagnetic to ferromagnetic (P-FM) transition that can be noticed for δ=0.50 in Figure 12, most probably corresponds to the evolution of the magnetic order of the Co ions with a prevalence of the FM component due to the canting of the AFM ordering. The narrow range of this transition (from 286 to 274 K) perfectly match the sudden rise of the electrical conductivity of the sample. This can be understood coupling the structural information and the electrical and magnetic measurements. At the temperature where the Co ions start to order magnetically and orbitally there is an abrupt change in the lattice parameters of the orthorhombic unit cell and in particular a strong contraction of the inter-octahedra Co-O-Co bond angle from about 176° to about 168°. It is clear that this change has a dramatic influence on the electron transfer integral leading to charge localization. Further cooling from 274 to 258 K leads to the evolution of the AFM structure with a magnetic moment nearly equal to zero, thus suggesting an effective long-range AFM order involving the whole structure and the absence of disorder in the oxygen network.

The sample with δ=0.45 is constituted by a minor orthorhombic phase (ca. 10%) and a major tetragonal phase, as explained in details in the x-ray diffraction section. It is interesting to note that, starting from high temperature, the first magnetic transition (P-FM-AFM) nicely overlaps with the one observed for δ=0.50 and has to be connected to the orthorhombic phase of the sample which most probably has an oxygen content slightly higher than 5.45. The second transition, related to the majority tetragonal phase, has, overall, the same features of the high-$T$ transition with a first quite narrow rise of the magnetization, followed by an AFM transition. The most notably difference is the $T$-extension of the FM phase which is from about 220 to 180 K. After the completion of the AFM ordering the net magnetization increases with respect to the δ=0.50 sample (~0.11 emu/mol for δ=0.50 and ~0.25 emu/mol for δ=0.45). The further reduction of the oxygen content to 5.39 leads to a broadening of the P-FM-AFM transition with wider intervals for the set-up of the FM and AFM states. Comparing the three samples it is clear that the δ reduction leads to a lowering of the net magnetic moment at the maximum of the P-FM transitions, to a broadening of the existence *T*-range of the FM component and to an increase of the magnetic moment after the completion of the AFM transition.

The first relevant change between the three samples lowering δ is the progressive reduction of the Co ions with the formation of $Co^{2+}$ species. These ions are believed to be in the HS state. For the $Co^{3+}$ ions we have estimated the spin-state from the susceptibility and *M vs. H* curves at different temperatures (not shown). At δ=0.50 the effective paramagnetic moment (above the IM transition) is around 8.8 which suggests the presence of almost all of the $Co^{3+}$ ions in the IS state. This agrees well with a dependence of the $Co^{3+}$ spin-state with the *RE* size: a recent report from Frontera *et al*. has shown that for the $YBaCo_2O_{5.5}$ sample (the Y and Ho ion sizes are 1.040 and 1.041 Å, respectively) the $Co^{3+}$ is in the IS spin state while by increasing the *RE* ion size part of the $Co^{3+}$ ions are in HS state. The IM transition observed at δ=0.5 is accompanied also by a spin-state transition of a fraction of $Co^{3+}$ IS to LS which has been found to be present in the AFM phase [24]. This spin configuration, the presence of all $Co^{3+}$ ions and the oxygen ordering assures a very effective and sudden set-up of the AFM state, as witnessed by the sharp FM-AFM transition.

For the samples with δ=0.45 and 0.395 we have again determined that most of the $Co^{3+}$ ions are in the IS state. However, at these oxygen content values it is also present $Co^{2+}$ in the HS state. According to the GK rules, it is predicted a weak ferromagnetic coupling between these two ions [17]. This couples well with the observation that the FM component seems to extend its *T*-range of existence along with the increase of HS $Co^{2+}$ from δ=0.50 to 0.395. However, beside this interaction other competing ones are the strong AF coupling between HS $Co^{2+}$ ions and the evolution, by lowering the temperature, of the AFM coupling between HS $Co^{2+}$ and IS $Co^{3+}$. Most probably, these interactions (particularly the first one) become dominant below δ=0.395 (i.e. by increasing the $Co^{2+}$ amount above 10%) thus leading to disappearance of any P-FM transition in the susceptibility curves (as a matter of fact a very broad bump above the paramagnetic curve of the $Ho^{3+}$ ion is still visible for the δ=0.31 sample).

At δ=0.223 and 0.008 the susceptibility curves do not reveal the presence of any relevant transition and they are representative of the main magnetic contribution coming from the paramagnetic $Ho^{3+}$ ions. A closer look to the $1/\chi_{mol}$ curves shows very small deviations from linearity in the range ~280-260 K. Since we have determined for both samples an IS for the $Co^{3+}$ ions these features may be due to a partial spin-transition to a LS state. However, up-to now, there is no information about the magnetic structure of intermediate δ compositions for the $HoBaCo_2O_{5+\delta}$ compound. With regard to this aspect low-temperature neutron diffraction measurements on the whole δ-range are already planned.

Finally, the δ=0 sample reveals the presence of the two already reported magnetic transitions, as observed by DSC measurements (see previously in the text). From the linear part of the $\chi_{mol}$ vs. $T$ curve we determined that all the $Co^{3+}$ ions are now in the HS state, in agreement with previous reports [17]. However, let us note from the $1/\chi_{mol}$ curve of this sample a clear transition from 288 to 264 K which closely resembles the P-FM-AFM transitions observed for higher δ-values. This feature does not fall in the range of the AFM ordering observed from DSC data (at about 340 K) nor in the region of the $T_{CO}$ (ca. 210 K). The $T$-range of this transition very nicely matches the $T$-interval of the I-M transition of the δ=0.50 sample where also a partial IS-LS transition occurs. The most probable explanation of this peculiarity for the δ=0 sample is a HS-IS transition of all or a fraction of the $Co^{3+}$ ions. Magnetic studies on the $HoBaCo_2O_5$ sample could not discriminate between the two probable models involving $Co^{3+}$ ions either in HS or in IS state. The observed G-type AFM structure does not provide additional information since both models may be qualitatively explained by the GK rules for superexchange. We believe that our experimental results strongly support a HS-IS transition due also to the development of a weak FM for a small temperature interval as a consequence of the coupling along the *b*-direction of the orthorhombic unit cell. As a further proof let us remember that all of the possible magnetic interactions along the three crystallographic directions between HS $Co^{3+}$ and HS $Co^{2+}$ are strongly AFM [26,27]. However, the ferromagnetic IS-$Co^{3+}$-HS-$Co^{2+}$ coupling is relatively weak and strongly dependent upon the Co-O-Co bond angle and its existence is only possible in a narrow range where the structural conditions are favorable.

By further increasing the oxygen content above δ=0.50 (i.e. in the mixed $Co^{3+}/Co^{4+}$ state) the HoBaCo$_2$O$_{5.55}$ sample has a magnetic transition which closely resembles the one found at δ=0.50 which, however, is shifted towards lower *T*-values and appears to be broader with respect to that sample, in perfect agreement with the results found for the HoBaCo$_2$O$_{5+δ}$ system [22]. In this case the $Co^{4+}$ ions are in the LS state while the estimation of the spin-state for the $Co^{3+}$ ions suggest that the ions are preferentially in the IS state. It is clear that the presence of a small amount of $Co^{4+}$ ions strongly influences the magnetic coupling of the $Co^{3+}$ ions by slightly shifting the P-FM transition to lower temperature and extending the existence *T*-interval of the FM state. Also in this case it can be expected that the FM phase results from a canted AFM structure which then evolves to a "normal" AF structure at lower temperature. From these results it seems that the $Co^{4+}$ ions have the influence to weaken the background AFM transition of the $Co^{3+}$ ions but at the concentration present in the δ=0.55 sample the effect is not detrimental for the set-up of the AFM state observed also in the HoBaCo$_2$O$_{5.5}$ composition.

Finally, it can be observed that the $θ_p$ value extracted from the slope of the $χ^{-1}$ curve at high temperature progressively evolves from a high negative value at δ=0 (ca. -100) to a positive value at δ=0.55 thus indicating the progressive strengthening of ferromagnetic interactions by increasing the oxygen content.

# Conclusion

In this work we have carried out the first systematic investigation of the role of oxygen content variation on the HoBaCo$_2$O$_{5+\delta}$ layered cobaltite. Up to now, the only available experimental information on this composition was related to the $\delta=0$ and 0.5 samples. The main conclusions of this work can be summarized in the following:

1 – We have determined, for the first time, the dependence with $T$ and $p(O_2)$ of the oxygen content for the HoBaCo$_2$O$_{5+\delta}$ layered cobaltite; experimental data have shown that under ambient pressure the maximum oxygen content achievable is 5.5 while for cobaltites containing larger rare-earths (*RE*) it is possible to have, at the same $T$ and $p(O_2)$ values, higher oxygen contents [9];

2 – At room temperature the HoBaCo$_2$O$_{5+\delta}$ crystal structure is orthorhombic (*Pmmm*) for $\delta=0$ with a doubling of the ideal perovskite unit cell along the *c* direction and a very small orthorhombic distortion; in the range 0<$\delta$<0.50 the unit cell is tetragonal (*P4/mmm*) with the doubling along the *c* direction; and finally for 0.50≤$\delta$≤0.55 the crystal structure is orthorhombic (*Pmmm*) with a doubling of the ideal perovskite unit cell along the *b* and *c* directions due to the oxygen vacancy ordering;

3 – The HoBaCo$_2$O$_{5+\delta}$ layered cobaltite shows a very strong and subtle dependence of its properties as a function of the oxygen content; in other members of the *RE*BaCo$_2$O$_{5+\delta}$ family, for example, I-M and magnetic transitions are observed in a wider range around the $\delta=0.50$ value. For example, when *RE*=Pr, the I-M transition is found up to $\delta=0.40$ while in the present case monophasic and oxygen homogeneous samples displaying a I-M transition are found only for $\delta\geq0.50$;

4 – The strong AFM interactions present at $\delta=0$ progressively weaken by increasing the oxygen content allowing also the set-up of a FM order as a consequence of Co oxidation and the change of the spin-state from HS Co$^{3+}$ at $\delta=0$ to IS for $\delta>0$.

5 – The mixed valence $Co^{3+}/Co^{4+}$ enhances the ferromagnetic interactions, as indicated by a positive $\theta_p$ value; the I-M and P-FM transitions shift at lower temperature passing from $\delta=0.50$ to $0.55$ as a consequence of the hole doping; however at $\delta=0.55$ the charge carriers are less delocalized in the metal-like phase as shown by a conductivity one order of magnitude lower with respect to $\delta=0.50$, most probably as a consequence of a smaller inter-polyhedra bond angle.

This work has shown, together with previous works on other $RE$BaCo$_2$O$_{5+\delta}$ systems, that the properties of layered cobaltites are influenced by a fine interplay between two major inter-connected degrees of freedom which are dominated by the oxygen content variation the Co oxidation state and $Co^{3+}$ spin-state. This in turn influences the structural properties of the system, the nature and strength of the magnetic coupling and the charge carrier nature and localization. As a matter of fact, it looks that a full comprehension of the layered cobaltites properties requires a thorough experimental multi-technique approach which should in any case start with the exact knowledge of the oxygen content as a function of temperature, oxygen partial pressure and thermal treatments.

# Acknowledgement

This work has been supported by the "Celle a combustibile ad elettroliti polimerici e ceramici: dimostrazione di sistemi e sviluppo di nuovi materiali" FISR Project of Italian MIUR. We recognize the support of the UNIPV-Regione Lombardia Project on Material Science and Biomedicine. The Authors are indebted with Prof. Maignan (CRISMAT) for performing the high-pressure oxygen treatment on the samples.

**Table 1** – Space group, lattice constants and cell volume for the various samples investigated in the present work. For the δ=0.45 sample we reported the data related to the major tetragonal phase of the sample.

|   | δ=0 | δ=0.01 | δ=0.02 | δ=0.22 | δ=0.31 | δ=0.39 | δ=0.45 | δ=0.50 | δ=0.55 |
|---|---|---|---|---|---|---|---|---|---|
| *Space Group* | *Pmmm* | *P4/mmm* | *P4/mmm* | *P4/mmm* | *P4/mmm* | *P4/mmm* | *P4/mmm* | *Pmmm* | *Pmmm* |
| *a* | 3.8975(1) | 3.8930(1) | 3.8892(1) | 3.87851(7) | 3.87766(7) | 3.87519(7) | 3.87380(7) | 3.85027(6) | 3.8271(1) |
| *b* | 3.8897(1) | 3.8930(1) | 3.8892(1) | 3.87851(7) | 3.87766(7) | 3.87519(7) | 3.87380(7) | 7.8209(1) | 7.8497(4) |
| *c* | 7.4761(2) | 7.4814(2) | 7.4854(2) | 7.4968(1) | 7.4966(1) | 7.4986(1) | 7.5009(2) | 7.5099(1) | 7.5231(4) |
| *V* | 113.390(3) | 113.383(4) | 113.226(3) | 112.773(3) | 112.721(3) | 112.608(3) | 112.561(4) | 113.074(3) | 113.003(3) |

# Figures

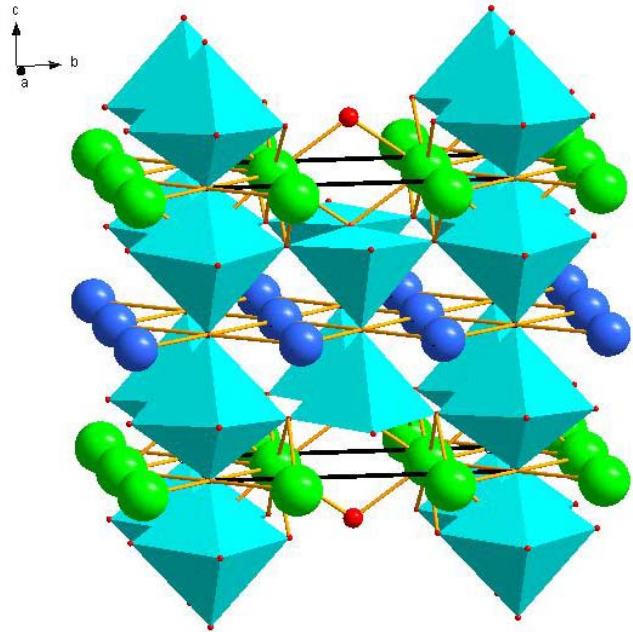

**Figure 1** – Sketch of the HoBaCo$_2$O$_{5.5}$ orthorhombic unit cell with a doubling along the *b*-direction. Light blue polyhedral represent the coordination of Co ions. Green spheres are Ba ions, the blue ones the Ho ions and the red ones the O ions.

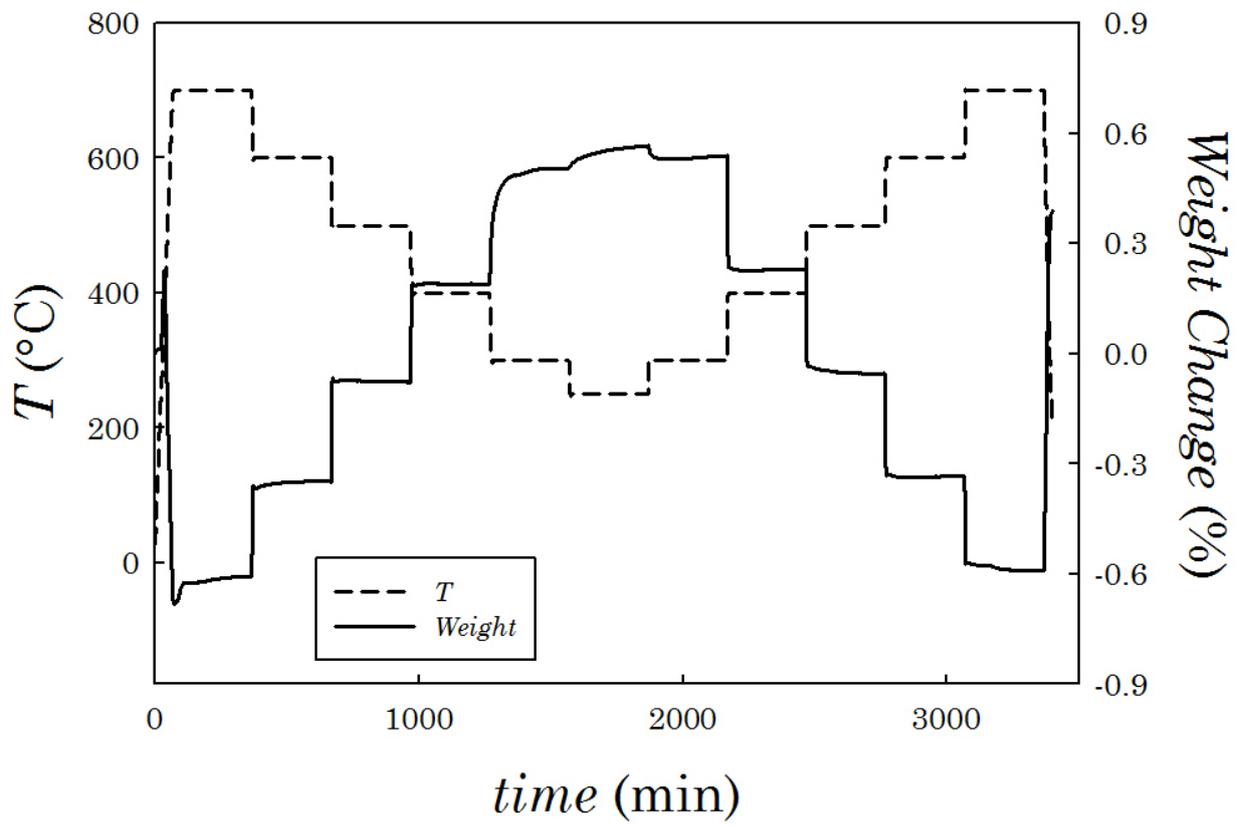

**Figure 2** – Weight change of $HoBaCo_2O_{5+\delta}$ as a function of temperature in pure oxygen.

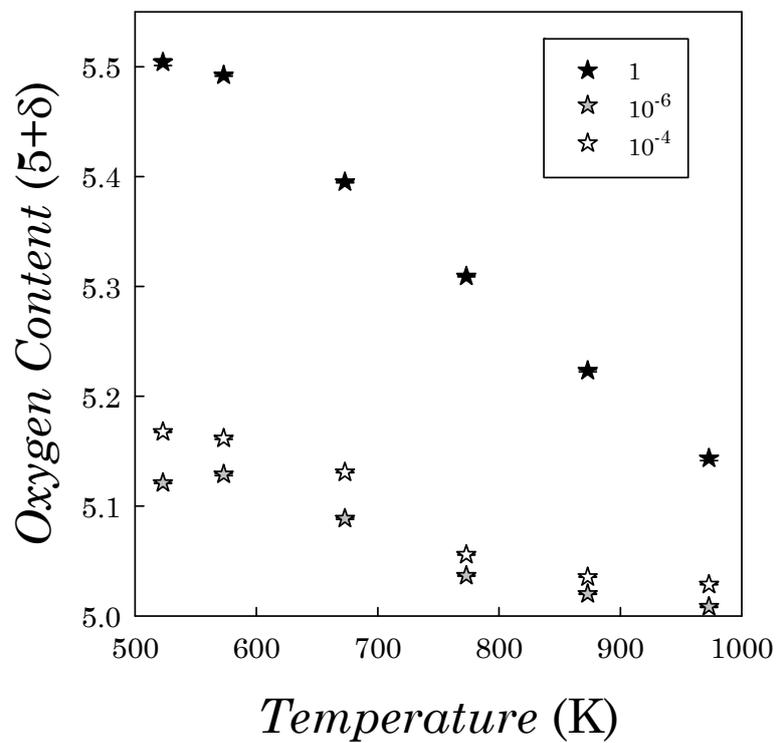

**Figure 3** – Oxygen content of HoBaCo$_2$O$_{5+\delta}$ as a function of temperature under various $p$(O$_2$).

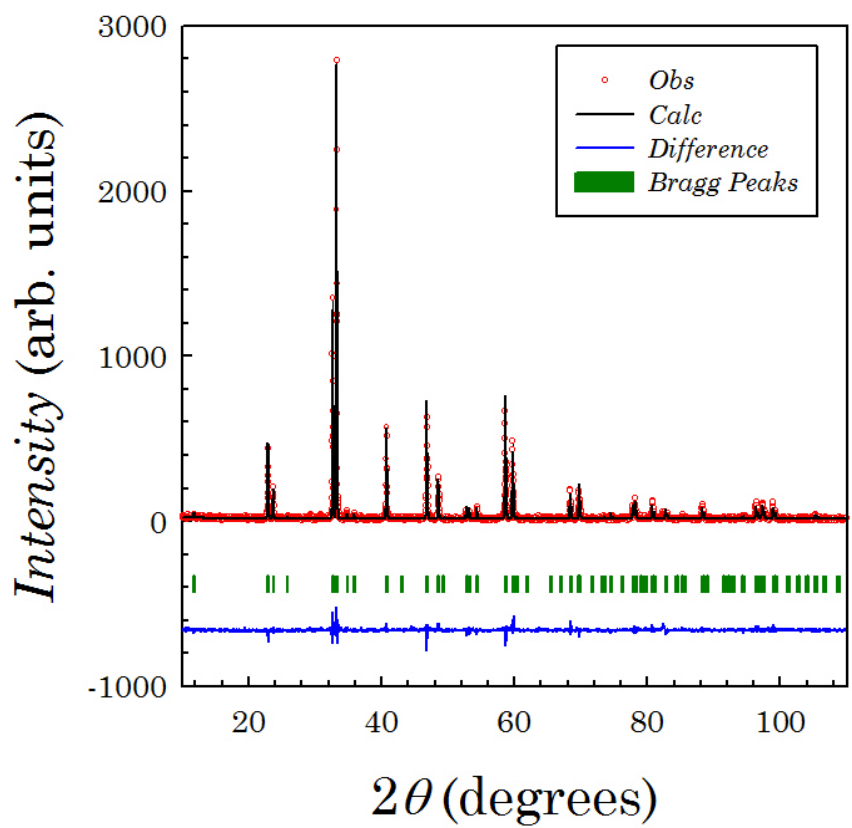

**Figure 4 -** Rietveld refined pattern of $HoBaCo_2O_{5+\delta}$. Red empty circles represent the experimental pattern, black line the calculated one while vertical green bars at the bottom of the pattern are the Bragg peaks positions. Horizontal blue line shows the difference between the calculated and experimental patterns.

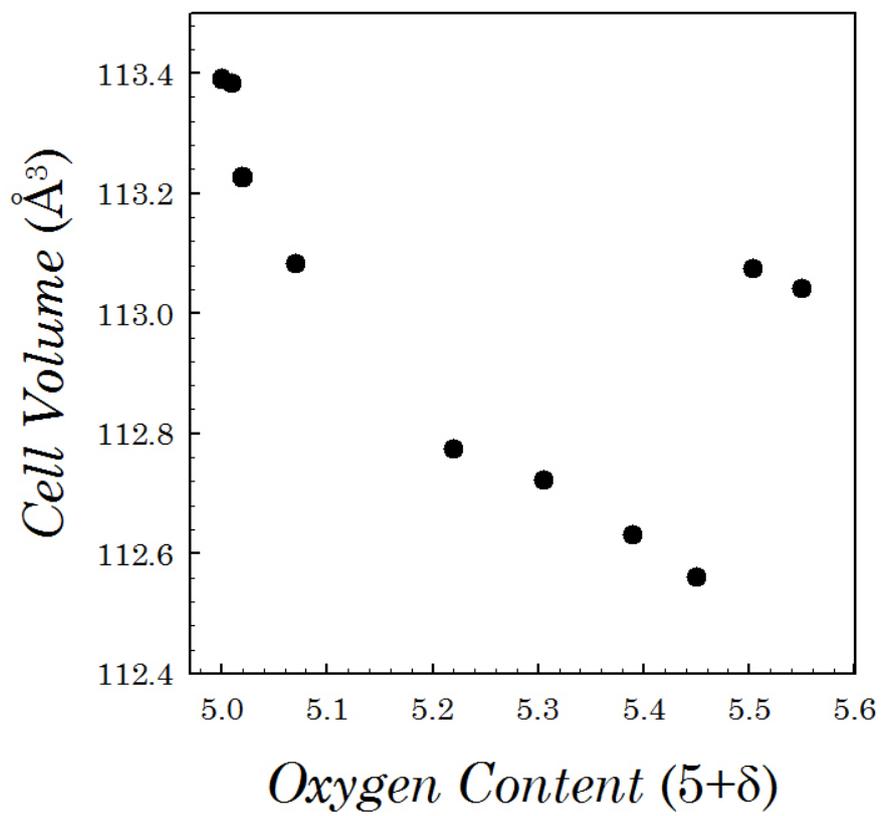

**Figure 5** — Cell volume as a function of δ for HoBaCo$_2$O$_{5+\delta}$.

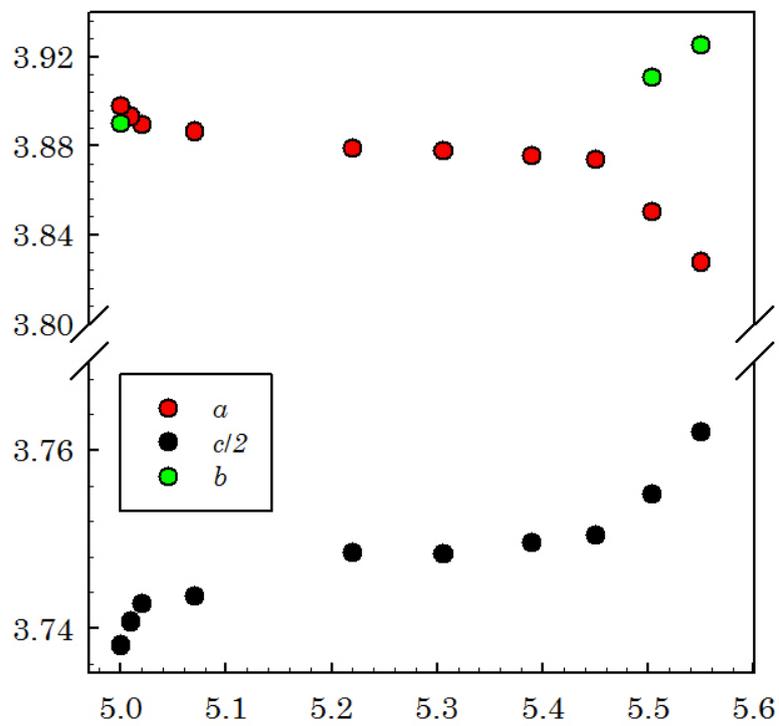

**Figure 6** — Lattice constants variation as a function of δ for HoBaCo$_2$O$_{5+\delta}$.

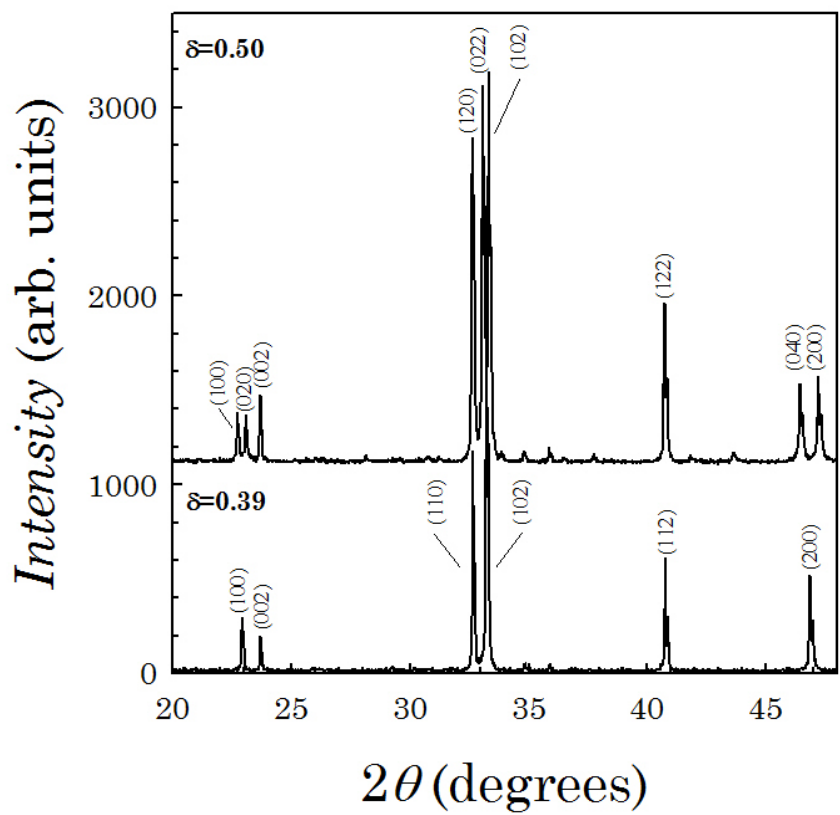

**Figure 7** － Indexed patterns for HoBaCo$_2$O$_{5+\delta}$ with $\delta$=0.39 and 0.50.

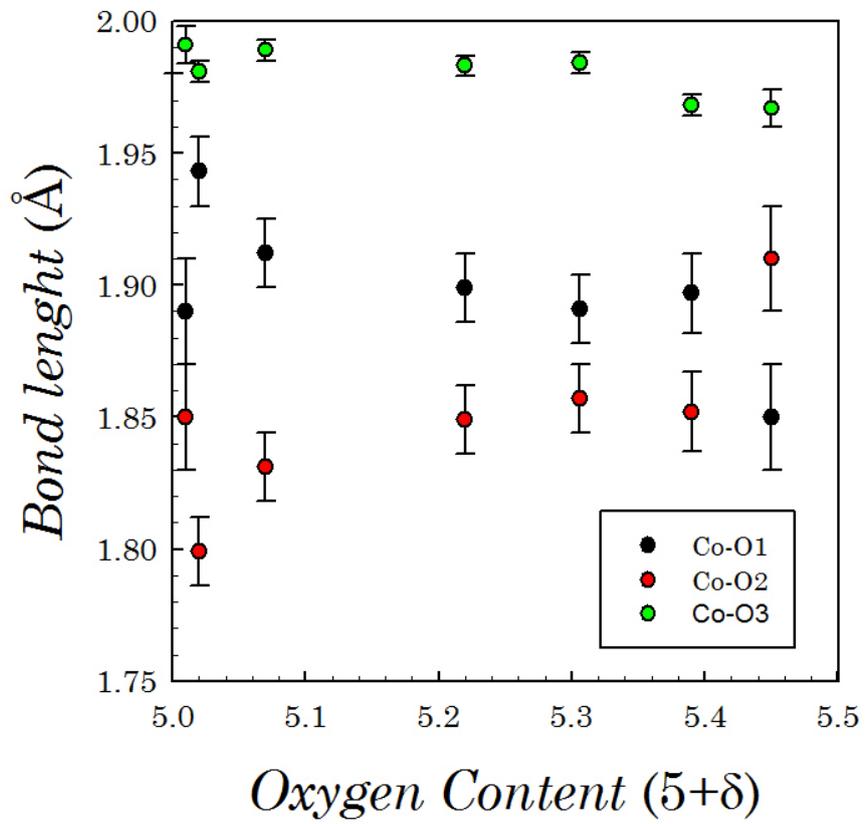

**Figure 8** － Bond length variation as a function of δ in the tetragonal phase of $HoBaCo_2O_{5+\delta}$.

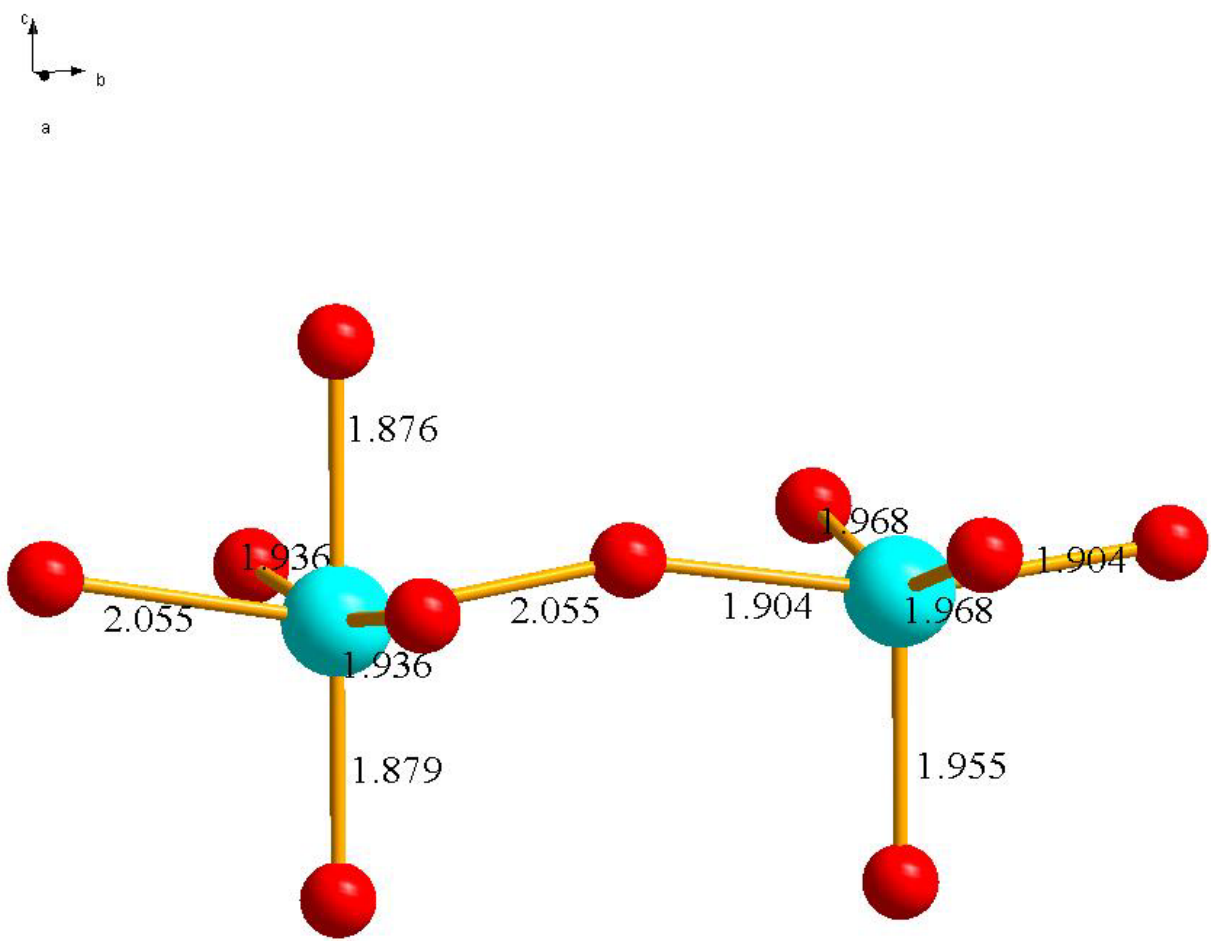

**Figure 9** — Bond lengths in the CoO5 and CoO6 polyhedra for the HoBaCo$_2$O$_{5.5}$ sample.

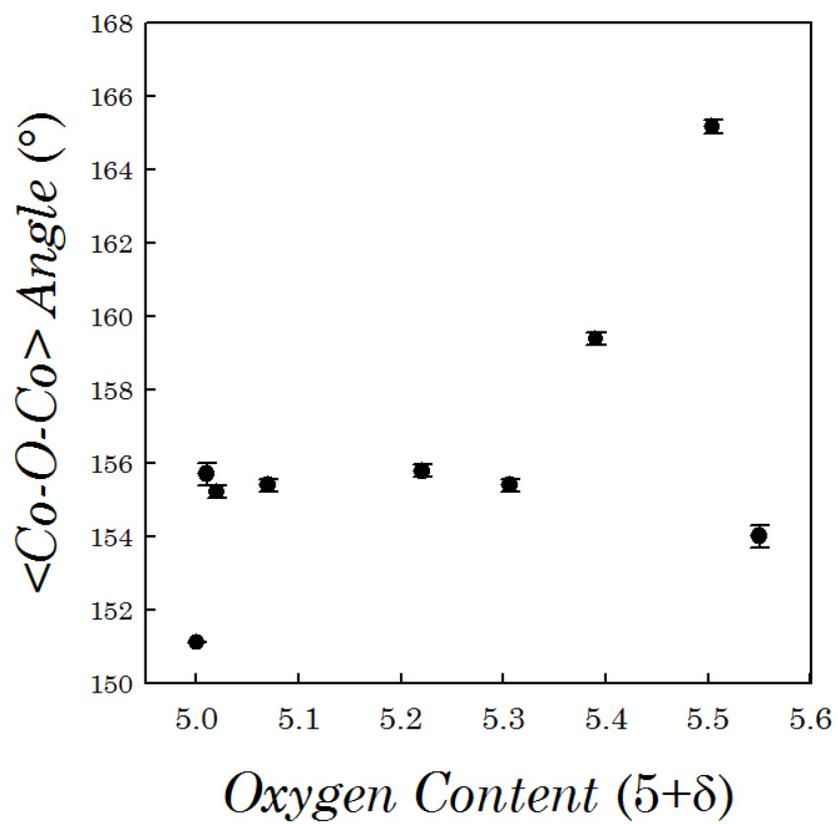

**Figure 10 −** In-plane Co-O-Co bond angle variation as a function of δ for $HoBaCo_2O_{5+\delta}$.

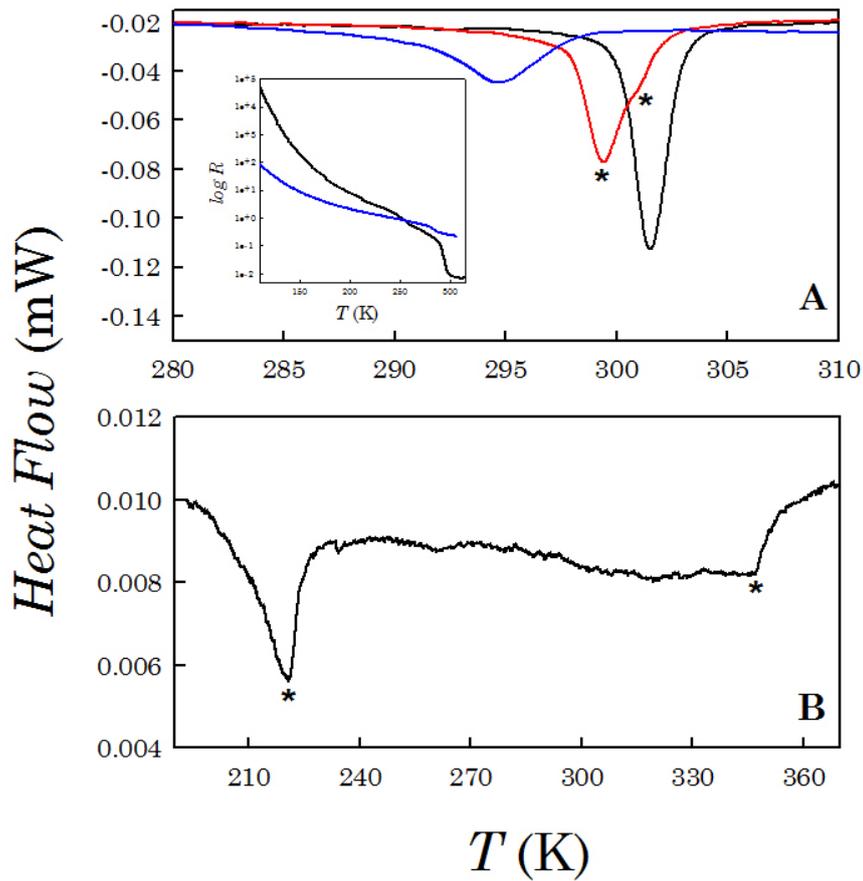

**Figure 11** — Panel A: DSC curves for HoBaCo$_2$O$_{5.5}$ and for another sample slowly cooled down to room temperature in pure oxygen from 700°C. In the inset: logarithm of electrical resistivity *vs*. *T* for HoBaCo$_2$O$_{5.5}$. Panel B: DSC curve for HoBaCo$_2$O$_5$.

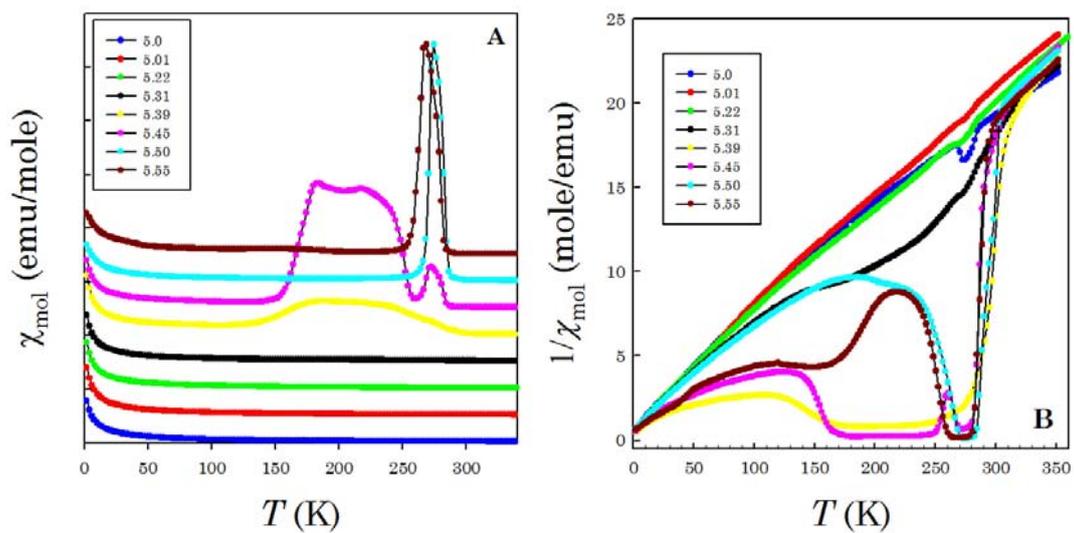

**Figure 12** ― Molar susceptibility (panel A - shifted for visualization purposes) and inverse molar susceptibility (panel B) of the HoBaCo$_2$O$_{5+\delta}$ samples investigated.